\documentclass[12pt]{article}

 \usepackage{amssymb}
 \usepackage{amsmath}
 \usepackage{amsfonts}
 \usepackage{color}
 \usepackage{dcolumn}
 \usepackage{graphicx}
 \numberwithin{equation}{section}

 \newcommand{\be}{\begin{equation}}
 \newcommand{\ee}{\end{equation}}
 \newcommand{\bea}{\begin{eqnarray}}
 \newcommand{\eea}{\end{eqnarray}}

\usepackage{cite}

\begin{document}

 \begin{titlepage}
  \thispagestyle{empty}

\vspace*{1mm}%
\hfill%
\vbox{
    \halign{#\hfil        \cr
           WITS-CTP-101\cr
          % SUT-P-07-2b   \cr
                     } % end of \halign
      }  % end of \vbox
\vspace*{15mm}%

 % \vspace{2cm}

  \begin{center}
    \font\titlerm=cmr10 scaled\magstep4
    \font\titlei=cmmi10 scaled\magstep4
    \font\titleis=cmmi7 scaled\magstep4

     \centerline{\titlerm  Nonplanar Integrability and Parity in ABJ Theory}

\vspace*{15mm} \vspace*{1mm} {\bf {\large Badr Awad Elseid Mohammed}}{$^{1,2}$}

 \vspace*{1cm}

{\it ${}^1$ National Institute for Theoretical Physics ,\\
Department of Physics and Centre for Theoretical Physics\\
University of Witwatersrand, Wits, 2050,\\
South Africa\\
${}^2$ Department of Physics,\\ Sudan University of Science and Technology\\
407, Sudan \\ }

\vspace*{2cm}

e-mail: bmohamme@ictp.it

\end{center}

\vspace*{2cm}

  \begin{abstract}
In this article we study the action of the non-planar two-loop dilatation operator in an {$SU\left(2\right)\times SU\left(2\right)$} sub-sector of the ABJ Chern-Simons-matter theory. The gauge invariant operators we consider are the restricted Schur polynomials. As in ABJM theory, there is a limit in which the spectrum reduces to a set of decoupled harmonic oscillators, indicating integrability in the large {$M$} and {$N$} double limit of the theory. We then consider parity transformations on the gauge invariant operators. In this case the non-planar anomalous dimensions break parity invariance. Our analysis shows that {$\left(M-N\right)$} is related to the holonomy in the string theory, confirming one of the main features of the theory and its string dual. Furthermore, in the limit where ABJ theory reduces to ABJM theory, parity invariance is restored.

  \end{abstract}

\end{titlepage}
\tableofcontents
\section{Introduction}
The AdS/CFT correspondence \cite{Maldacena} claims a duality between string theory or M-theory on the negatively curved background of anti de Sitter space in {$\left(d+1\right)$}-dimensions ({$AdS_{d+1}$} ) and a conformal field theory living on the boundary of this space, which is d-dimensional Minkowski space. The first example of such a correpondence is the duality between {$\mathcal{N}=4$} super-Yang-Mills (SYM) theory in 4-dimensions and type IIB superstring theory on {$AdS_{5}\times S^{5}$}. Recently, new dualities were found, in which {$\mathcal{N}=6$} superconformal Chern-Simons-matter theory in 3-dimensions is dual to type IIA superstring theory on {$AdS_{4}\times CP^{3}$}. The first example involves ABJM theory with gauge group {$U\left(N\right)_{k}\times\overline{U\left(N\right)}_{-k}$}, where {$k$} denotes the Chern-Simons level \cite{ABJM}. This was then generalized to gauge group {$U\left(M\right)_{k}\times\overline{U\left(N\right)}_{-k}$} giving the ABJ theory \cite{ABJ}. One consequence of these dualities is a relation between the energy of string states and the conformal dimension of the dual operators. The conformal dimension is given by the eigenvalue of the dilatation operator of the CFT side. Therefore a good test of these dualities is to compute the action of the dilatation operator on a gauge invariant operator in the CFT and compare this to energies of states in the string theory. Further, there are good reasons to hope that the dilatation operator maps onto the Hamiltonian of an integrable system \cite{Bethe,spin-chain}.\\

 A system is integrable if it has one conserved quantity for each degree of freedom. It allows one to find the exact spectrum of the dilatation operator. The discovery of integrability in {$\mathcal{N}=4$} SYM and ABJM theories \cite{Bethe,dilatation,spin-chain} in the planar limit has allowed dramatic progress. It is thus natural to look for integrability in the non-planar limit. Non-planar integrability may be important for a study of the quantum dynamics of string theory.\\

  A new technique employing the representation theory of symmetric and unitary groups \cite{Schur-polynomials0,Schur-polynomials1,Schur-polynomials2} has been used to compute the dilatation operator of {$\mathcal{N}=4$} super-Yang-Mills theory on restricted Schur polynomials in a large {$N$} but non-planar limit at one loop \cite{nonplanar1,nonplanar2,nonplanar3,nonplanar4} and at two loops \cite{nonplanar-twoloop}. The results of this work demonstrated integrability in a large {$N$} but non-planar limit. Motivated by these results, the action of the two loop dilatation operator of ABJM theory in its {$SU\left(2\right)\times SU\left(2\right)$} sector was studied \cite{nonplanar-ABJM} suggesting the theory enjoys non-planar integrability at two loops. In contrast to this, the spin chains technique for ABJM which is relevant for expansion about the planar limit of the theory did not find any sign of non-planar integrability \cite{nonplanar-ABJM2}. In \cite{nonplanar-ABJM2,nonplanar-ABJ} the action of the non-planar dilatation operator has been studied on operators with charge $O(1)$. Furthermore, the non-planar diagrams summed in \cite{nonplanar-ABJM2,nonplanar-ABJ} are correcting the large $N$ limit, since they have been considered as a correction to the leading planar diagrams using quantum mechanical perturbation theory. The non-planar diagrams we sum are very different to \cite{nonplanar-ABJM2,nonplanar-ABJ}, since we study the action of the non-planar dilatation operator on operators with charge $O(N)$. In this case the non-planar diagrams we sum contribute to the leading term in the large {$N$} limit.\\

ABJ theory manifests parity violation in the supersymmetric gauge theory and its dual string theory. In the field theory side, parity takes {$U\left(M\right)_{k}\times\overline{U\left(N\right)}_{-k}$} to {$U\left(N\right)_{k}\times\overline{U\left(M\right)}_{-k}$} while in string theory side parity violation is encoded into the non-trivial holonomy {$b_{2}$} on {$CP^{1}\subset CP^{3}$} with a background NS {$B$}-field {${\cal{B}}_{2}$} given by
\begin{equation}\label{holonomy}
b_{2}\equiv\frac{1}{2\pi}\int_{CP^{1}\subset CP^{3}}{\cal{B}}_{2}=\frac{M-N}{k}.
\end{equation}

In contrast to ABJM theory which has constant holonomy, ABJ theory has {$(M-N)$} holonomy dependence in its string dual\footnote{More precisely, ABJM has holonomy {$b_2=\frac{1}{2}$} while ABJ has {$b_2=\frac{1}{2}+\frac{M-N}{k}$}.}. This holonomy breaks parity for {$M\neq N $}. Therefore, to test the conjectured duality in ABJ theory, it is interesting to examine the spectrum of operators in this theory in the double 't Hooft limit and to check whether {$\left(M-N\right)$} can be related to a {$B$} field. This test has been done in \cite{nonplanar-ABJ} using the spin chain representation.\\

It has been shown in \cite{D.Bak,J.A.Minahan} that ABJ theory might still enjoy integrability in the presence of parity symmetry breaking. Determining if ABJ theory enjoys integrability in a large $N$ but non-planar limit is an open question.\\

The importance of this work is not only in the hints of integrability in the non-planar limit, but also it proves for the first time the co-existence of non-planar integrability and parity breaking effect as predicted in \cite{D.Bak,J.A.Minahan}.\\

In this paper we study the two loop non-planar anomalous dimensions of restricted Schur polynomials in the ABJ theory. In the large $N$ and $M$ double limit, the spectrum of the anomalous dimension reduced to a set of decoupled harmonics oscillator indicating non-planar integrability and exhibits the expected parity-breaking effect. In section 2, we compute the action of non-planar two loop ABJ dilatation operator on restricted Schur polynomials.  We then analyse the spectrum of the anomalous dimensions in section 3. The parity operation is considered in section 4. Finally section 5 contains our conclusion.
%%%%%%%%%%%%%%%%%%%%%%%%%%%%%%%%%%%%%%%%%%%%%%
%%%%%%%%%%%%%%%%%%%%%%%%%%%%%%%%%%%%%%%%%%%%%%
\section{ABJ Dilatation Operator}
%%%%%%%%%%%%%%%%%%%%%%%%%%%%%%%%%%%%%%%%%%%%%%
%%%%%%%%%%%%%%%%%%%%%%%%%%%%%%%%%%%%%%%%%%%%%%
ABJ theory is a  three-dimensional $\mathcal{N}=6$ superconformal Chern-Simons-matter theory with gauge group {$U\left(M\right)_{k}\times\overline{U\left(N\right)}_{-k}$ } and R-symmetry group {$SU\left(4\right).$} In this case the bifundamental scalar fields  {$A$} and {$B$} transform as
\begin{equation*}
A\rightarrow A^{\prime}=U\left(M\right)A\overline{U\left(N\right)}^{\dagger},
\end{equation*}
\begin{equation*}
B\rightarrow B^{\prime}=U\left(M\right)B\overline{U\left(N\right)}^{\dagger}.
\end{equation*}
 Therefore all traces constructed from the pairs {$AB^{\dagger}$} are invariant under {$U\left(M\right)_{k}\times\overline{U\left(N\right)}_{-k}$} gauge transformations.\\
The ABJ dilatation operator is closely related to the ABJM one, since both of them come from the F-terms of the bosonic potential. The key difference is in their coupling constants. In ABJM theory the coupling constant is $\left(\frac{\lambda}{N}\right)^{2}$ while in ABJ theory it is {$\left(\frac{\lambda}{N}\right)\left(\frac{\hat{\lambda}}{M}\right)$} where  {$\lambda=\frac{4\pi N}{k}$ and $\hat{\lambda}=\frac{4\pi M}{k} $} . Therefore in ABJ theory we have a double 't Hooft limit that is given by
\begin{equation*}
N,M\rightarrow\infty,\quad k\rightarrow\infty,\quad\lambda,\hat{\lambda}\:\mathrm{fixed.}
\end{equation*}
The dilatation operator of ABJ theory has a closed action on the {$SU(2)\times SU(2)$} subsector built with one type of excitation field {$B_{2}$} \cite{nonplanar-ABJ}. On this sector the dilatation operator is
\begin{equation}\label{dilatation}
D=\left(V_{F}^{bos}\right)^{eff}=-\frac{\lambda}{N}\frac{\hat{\lambda}}{M}:\mathrm{Tr\biggl[\left(B_{2}^{\dagger}A_{1}B_{1}^{\dagger}-B_{1}^{\dagger}A_{1}B_{2}^{\dagger}\right)\left(\frac{\partial}{\partial B_{2}^{\dagger}}\frac{\partial}{\partial A_{1}}\frac{\partial}{\partial B_{1}^{\dagger}}-\frac{\partial}{\partial B_{1}^{\dagger}}\frac{\partial}{\partial A_{1}}\frac{\partial}{\partial B_{2}^{\dagger}}\right)\biggr]}:,
\end{equation}
where {$A_{1}$}  and {$B_{1}^{\dagger}$} are the ``background'' fields, and {$:\space:$} means that all the fields in \eqref{dilatation} should not be self-contracted. \\
Like the ABJM case, we will study the action of dilatation operator \eqref{dilatation} on the gauge invariant operators built from restricted Schur polynomials for the gauge group {$U\left(M\right)_{k}\times\overline{U\left(N\right)}_{-k}$} . With one type of ``excitation'' {$B_{2}^{\dagger} $} these polynomials are
\begin{equation}\label{operator}
O_{R,\left\{ r\right\} }=\frac{1}{\prod_{ij}n_{ij}!}\sum_{\sigma\in S_{n}}\mathrm{Tr}_{\left\{ r\right\} }\left(\Gamma_{R}\left(\sigma\right)\right)\prod_{j=1}^{m_{2}}\left(A_{1}B_{2}^{\dagger}\right)_{a_{\sigma\left(j\right)}}^{a_{j}}\prod_{i=m_{2}+1}^{n}\left(A_{1}B_{1}^{\dagger}\right)_{a_{\sigma\left(i\right)}}^{a_{i}},
\end{equation}
where the label {$R$} specifies an irreducible representation ``{$irrep$}'' of the symmetric group {$S_{n}$} and {$\left\{ r\right\} \equiv\left\{ r_{12},r_{11}\right\} $} is an irreducible representation of {$S_{n_{12}}\times S_{n_{11}}\subset S_{n}$}. We have {$n_{12}+n_{11}=n$}.\\
In this expression {$m_{2}=n_{12}$} is the number of {$A_{1}B_{2}^{\dagger}$} pairs and {$m_{1}=n_{11}$} is the number of {$A_{1}B_{1}^{\dagger}$} pairs. We will study the limit in which the number of background fields is much bigger than the number of excitation fields, that is, {$m_{2}\ll m_{1}$}.\\
From appendix A, the bare two point function of operators with one type of excitation fields {$B_{2}$} is
\begin{equation*}
\left\langle O_{R,\left\{ r\right\} }O_{S,\left\{ s\right\} }^{\dagger}\right\rangle =\delta_{RS}\delta_{\left\{ r\right\} ,\left\{ s\right\} }\frac{\mathrm{hooks_{R}}f_{R}(M)f_{R}(N)}{\mathrm{hooks_{\mathrm{}}}_{r_{11}}\mathrm{hooks}_{r_{12}}}.
\end{equation*}
In this case, the operator {$O_{R,\left\{ r\right\} }$} is related to the normalized operator {$\widehat{O}_{R,\left\{ r\right\} }$} as
\begin{equation*}
O_{R,\left\{ r\right\} }=\sqrt{\frac{\mathrm{hooks_{R}}f_{R}(M)f_{R}(N)}{\mathrm{hooks_{\mathrm{}}}_{r_{11}}\mathrm{hooks}_{r_{12}}}}\hat{O}_{R,\left\{ r\right\} }.
\end{equation*}
Before we study the action of dilatation operator on \eqref{operator}, it is important to point out that in ABJ theory both {$A$} and {$B$} are  matrix fields, and hence we have the following index structure
\begin{equation}
\left(A_{k}B_{l}^{\dagger}\right)_{a_{\sigma\left(i\right)}}^{a_{i}}=\left(A_{k}\right)_{\alpha}^{a_{i}}\left(B_{l}^{\dagger}\right)_{a_{\sigma\left(i\right)}}^{\alpha},\qquad\alpha=1,2,...,M,
\end{equation}
where {$k,\, l\in\left\{ 1,2\right\} $}. The result of acting with \eqref{dilatation} on \eqref{operator} is thus
\begin{equation}\label{action}
D\hat{O}_{R,\left\{ r\right\} }=\sum_{S,\left\{ s\right\}}M_{R,\left\{ r\right\} ;S,\left\{ s\right\} }\hat{O}_{S,\left\{ s\right\} }
\end{equation}
 where
 \begin{align}
 &M_{R,\left\{ r\right\};S,\left\{ s\right\} }=\sqrt{\frac{\textrm{hooks}_{S}f_{S}(M)f_{S}(N)\textrm{hooks}_{r_{11}}\textrm{hooks}_{r_{12}}}{\textrm{hooks}_{R}f_{R}(M)f_{R}(N)\textrm{hooks}_{s_{11}}\textrm{hooks}_{s_{12}}}}\left(\frac{4\pi}{k}\right)^{2}\sum_{R^{\prime},\,S^{\prime}}\frac{m_{1}m_{2}c_{RR^{\prime}}(M)d_{S}}{d_{s_{11}}d_{s_{12}}nd_{R^{\prime}}}\times\nonumber\\
 &\biggl[M\mathrm{Tr}\left(I_{S^{\prime}R^{\prime}}\left[\Gamma_{R}\left(\left(1,m_{2}+1\right)\right),P_{R,\left\{ s\right\} }\right]I_{R^{\prime}S^{\prime}}C\right)+\nonumber\\
  &+\mathrm{Tr}\left(I_{S^{\prime}R^{\prime}}\left[\Gamma_{R}\left(\left(1,m_{2}+1\right)\right)P_{R,\left\{ s\right\} }\Gamma_{R}\left(\left(1,m_{2}+1\right)\right)-P_{R,\left\{ s\right\} }\right]I_{R^{\prime}S^{\prime}}C\right)+\nonumber\\
 &+\left(m_{1}-1\right)\mathrm{Tr}\left(I_{S^{\prime}R^{\prime}}\left[\Gamma_{R}\left(\left(1,m_{2}+2\right)\right)P_{R,\left\{ s\right\} }\Gamma_{R}\left(\left(1,m_{2}+2\right)\right)\right]I_{R^{\prime}S^{\prime}}C\right)\nonumber\\
 &-\left(m_{1}-1\right)\mathrm{Tr}\left(I_{S^{\prime}R^{\prime}}\left[\Gamma_{S}\left(\left(m_{2}+2,1,m_{2}+1\right)\right)P_{R,\left\{ s\right\} }\right]I_{R^{\prime}S^{\prime}}C\right)+\nonumber\\
 &+\left(m_{2}-1\right)\mathrm{Tr}\left(I_{S^{\prime}R^{\prime}}\left[\Gamma_{R}\left(\left(1,m_{2}+1\right)\right)P_{R,\left\{ s\right\} }\Gamma_{R}\left(\left(1,2\right)\right)-P_{R,\left\{ s\right\} }\Gamma_{S}\left(\left(2,1,m_{2}+1\right)\right)\right]I_{R^{\prime}S^{\prime}}C\right)\biggr]\nonumber
 \end{align}
and
\begin{equation*}
C=\left[P_{R,\left\{ s\right\} },\Gamma_{S}\left(\left(1,m_{2}+1\right)\right)\right].
\end{equation*}
In this expression, $c_{RR^{\prime}}(M)$ is the weight of the removed box from the irrep $R$ to obtain $R^{\prime}$ \cite{nonplanar1,nonplanar2,Koch:2007uu}. $I_{R^{\prime}S^{\prime}}$ and $I_{S^{\prime}R^{\prime}}$ are the intertwiners defined in appendix B of \cite{nonplanar1}.\\
Let us pause at this point to discuss how our result \eqref{action} differs from the result for ABJM theory obtained in \cite{nonplanar-ABJM}. Firstly, \eqref{action} depends on $M$ and $N$ which allows us to consider interesting multiple scaling limits, which could in principle scale $M$ and $N$ independently. Secondly, the factor $c_{RR^{\prime}}(M)$ depends on $M$ while in the square root we have $(M,\,N)$ dependence. The action of the dilatation operator is clearly not symmetric upon interchanging $M\leftrightarrow N$.
%%%%%%%%%%%%%%%%%%%%%%%%%%%%%%%%%%%%%%%%%%%%%%%%%
%%%%%%%%%%%%%%%%%%%%%%%%%%%%%%%%%%%%%%%%%%%%%%%%%
\section{Spectrum of Anomalous Dimensions}
\begin{figure}[h]
  \hskip 3cm
  \includegraphics[width=0.8\textwidth,natwidth=600,natheight=25]{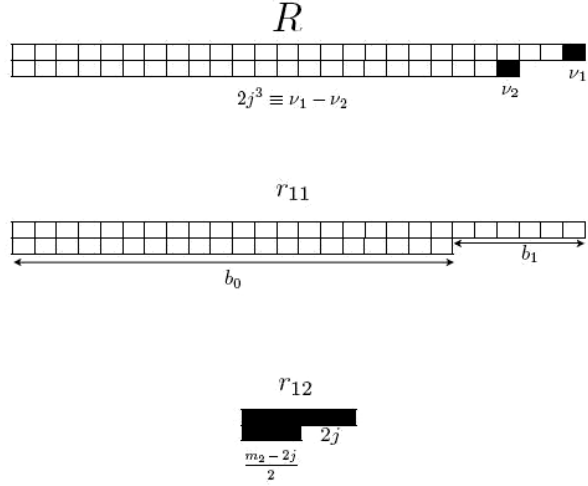}
  \caption{Construction of $r_{11}$ and $r_{12}$ from the irrep $R$.}\label{construction}
\end{figure}

%%%%%%%%%%%%%%%%%%%%%%%%%%%%%%%%%%%%%%%%%%%%%%%%%
%%%%%%%%%%%%%%%%%%%%%%%%%%%%%%%%%%%%%%%%%%%%%%%%%
To obtain the spectrum of the anomalous dimension from \eqref{action} we consider the case where the gauge invariant operators are labeled by Young diagrams of two long rows. Recall that we are interested in operators that have a bare dimension of order $N$. It has been shown in \cite{Schur-polynomials0,Koch:2007uu} that operators of dimension $O(N)$ are dual to giant gravitons. In this class of operators, one can construct the projectors using $U(2)$ group theory \cite{nonplanar2}. The irrep {$r_{11}$} is obtained from {$R$} by removing {$\nu_1$} and {$\nu_2$} boxes from the first and the second row of {$R$} respectively. The removed boxes {$\nu_1$} and {$\nu_2$} must be assembled into the irrep {$r_{12}$} with {$\nu_1+\nu_2=m_2$}. Figure \ref{construction} illustrates the construction of $r_{11}$ and $r_{12}$ from the irrep $R$. It is enough to specify {$\nu_1-\nu_2\equiv 2j^3$}. Specify the irrep {$r_{11}$} by {$b_{0}$} and {$b_{1}$} with {$2b_{0}+b_{1}=m_{1}$} where we consider {$m_{1}\gg m_{2}$}. Denote the number of boxes in the first row of {$r_{12}$} minus the number of boxes in the second row by {$2j$}. The number of boxes in the second row of {$r_{12}$} is thus {$\frac{m_2-2j}{2}$}.
\vskip 7cm
  The computation of \eqref{action} gives
\begin{equation*}
D\hat{O}_{j,j^{3}}\left(b_{0},b_{1}\right)=\left(\frac{4\pi}{k}\right)^{2}\biggl[\biggl(-\frac{M}{2}\left(m_{2}-\frac{\left(m_{2}+2\right)\left(j^{3}\right)^{2}}{j\left(j+1\right)}\right)
\end{equation*}
\begin{equation*}
-\frac{m_{2}^{2}}{4}+m_{2}+j_{3}^{2}-j\left(j+1\right)-\frac{j_{3}^{2}\left(4-m_{2}^{2}\right)^{2}}{4j\left(j+1\right)}\biggr)\triangle\hat{O}_{j,j^{3}}\left(b_{0},b_{1}\right)
\end{equation*}
\begin{equation*}
+M\sqrt{\frac{\left(m_{2}+2j+4\right)\left(m_{2}-2j\right)}{\left(2j+1\right)\left(2j+3\right)}}\frac{\left(j+j^{3}+1\right)\left(j-j^{3}+1\right)}{2\left(j+1\right)}\left(1+\frac{m_{2}-2j-4}{2N}\right)\triangle\hat{O}_{j+1,j^{3}}\left(b_{0},b_{1}\right)
\end{equation*}
\begin{equation}\label{spectrum}
+M\sqrt{\frac{\left(m_{2}+2j+2\right)\left(m_{2}-2j+2\right)}{\left(2j+1\right)\left(2j+3\right)}}\frac{\left(j+j^{3}\right)\left(j-j^{3}\right)}{2j}\left(1+\frac{m_{2}-2j-2}{2N}\right)\triangle\hat{O}_{j-1,j^{3}}\left(b_{0},b_{1}\right)\biggr],
\end{equation}
 where
\begin{align}\label{suboperator}
\triangle \hat{O}_{j,j^{3}}\left(b_{0},b_{1}\right)=&\sqrt{\left(M+b_{0}\right)\left(M+b_{0}+b_{1}\right)}\sqrt{\frac{\left(N+b_{0}+b_{1}\right)}{\left(N+b_{0}\right)}}\hat{O}_{j,j^{3}}\left(b_{0}-1,b_{1}+2\right)\nonumber\\
+&\sqrt{\left(M+b_{0}\right)\left(M+b_{0}+b_{1}\right)}\sqrt{\frac{\left(N+b_{0}\right)}{\left(N+b_{0}+b_{1}\right)}}\hat{O}_{j,j^{3}}\left(b_{0}+1,b_{1}-2\right)\nonumber\\
-&\left(2M+2b_{0}+b_{1}\right)\hat{O}_{j,j^3}\left(b_{0},b_{1}\right),
\end{align}
The combination {$\triangle \hat{O}_{j,j^{3}}\left(b_{0},b_{1}\right)$} in \eqref{suboperator} differs from the ones obtained in the case of  {${\cal{N}}=4$} SYM theory and ABJM theory \cite{nonplanar2,nonplanar-ABJM}, since it depends on both $M$ and $N$. One might also note that \eqref{spectrum} and \eqref{suboperator} are not invariant under the exchange $M\leftrightarrow N$. Note the {$j$} dependence in \eqref{suboperator} which implies the problem of diagonalizing the dilatation operator factorizes into two separate eigen problems. We also note that the spectrum of anomalous dimension in \eqref{spectrum} is different from the result obtained in \cite{nonplanar2}, therefore it is not clear how to obtain a direct diagonalisation  of \eqref{spectrum}. However, in the double limit
\begin{equation*}
M,\, N\rightarrow\infty,\qquad\frac{m_{2}}{M},\frac{m_{2}}{N}\ll1,
\end{equation*}
the leading contribution in \eqref{spectrum} is thus
\begin{equation*}
D\hat{O}_{j,j^{3}}\left(b_{0},b_{1}\right)=\left(\frac{4\pi}{k}\right)^{2}\biggl[\biggl(-\frac{M}{2}\left(m_{2}-\frac{\left(m_{2}+2\right)\left(j^{3}\right)^{2}}{j\left(j+1\right)}\right)\biggr)\triangle\hat{O}^{\prime}_{j,j^{3}}\left(b_{0},b_{1}\right)
\end{equation*}
\begin{equation*}
+M\sqrt{\frac{\left(m_{2}+2j+4\right)\left(m_{2}-2j\right)}{\left(2j+1\right)\left(2j+3\right)}}\frac{\left(j+j^{3}+1\right)\left(j-j^{3}+1\right)}{2\left(j+1\right)}\triangle\hat{O}^{\prime}_{j+1,j^{3}}\left(b_{0},b_{1}\right)
\end{equation*}
\begin{equation}\label{spectrum1}
+M\sqrt{\frac{\left(m_{2}+2j+2\right)\left(m_{2}-2j+2\right)}{\left(2j+1\right)\left(2j+3\right)}}\frac{\left(j+j^{3}\right)\left(j-j^{3}\right)}{2j}\triangle\hat{O}^{\prime}_{j-1,j^{3}}\left(b_{0},b_{1}\right)\biggr],
\end{equation}
where
\begin{eqnarray}
\Delta \hat{O}^{\prime}_{j,j^3}(b_{0},b_{1}) &&=\sqrt{(M+b_0)(M+b_0+b_1)}(\hat{O}_{j,j^3}(b_0+1,b_1-2)+\hat{O}_{j,j^3}(b_0-1,b_1+2))\nonumber \\
 &&-(2M+2b_0+b_1)\hat{O}_{j,j^3}(b_0,b_1).\label{DeltaDefn}
\end{eqnarray}
It is interesting to note that the result \eqref{spectrum1} is similar to the result of {$\mathcal{N}=4$} SYM theory multiplied by {$M$} \cite{nonplanar2}. In this context the continuum limit in ABJ theory reduces to a set of decoupled harmonic oscillators studied in \cite{nonplanar-ABJM}. Since the spectrum of harmonic oscillator can be easily determined, this indicates the integrability of ABJ theory in the large but non-planar double $M$ and $N$ limit. The spectrum of the anomalous dimension of ABJ theory is thus reduced to a set of oscillators. For $m_{2}=2n$, the frequency {$\omega_{i}$} and degeneracy {$d_{i}$} are given by
\begin{equation}\label{no-parity1}
\omega_{i}=8iM\left(\frac{4\pi}{k}\right)^{2},\quad d_{i}=2\left(n-i\right)+1,\quad i=0,1,...,n.
\end{equation}
For $m_{2}=2n+1$, the frequency {$\omega_{i}$} and degeneracy {$d_{i}$} are given by
\begin{equation}\label{no-parity2}
\omega_{i}=8iM\left(\frac{4\pi}{k}\right)^{2},\quad d_{i}=2\left(n-i+1\right),\quad i=0,1,...,n.
\end{equation}

%%%%%%%%%%%%%%%%%%%%%%%%%%%%%%%%%%%%%%%%
%%%%%%%%%%%%%%%%%%%%%%%%%%%%%%%%%%%%%%%%
\section{Parity Operation}
%%%%%%%%%%%%%%%%%%%%%%%%%%%%%%%%%%%%%%%%
%%%%%%%%%%%%%%%%%%%%%%%%%%%%%%%%%%%%%%%%
The action of the parity operator introduced in \cite{nonplanar-ABJ}, on the trace operator, inverts the order of the fields inside each of its traces, i.e.
\begin{equation*}
\hat{P}\,:\:\mathrm{Tr}\left(A^{a_{1}}B_{b_{1}}...A^{a_{l}}B_{b_{l}}\right)\rightarrow\mathrm{Tr}\left(B_{b_{l}}A^{a_{l}}...B_{b_{1}}A^{a_{1}}\right),
\end{equation*}
where {$a_i,b_i\in{1,2}$}. In this way, acting with the parity operation on the restricted Schur polynomial \eqref{operator} leads to
\begin{equation}\label{parity}
\hat{P}O_{R,\left\{ r\right\} }=\frac{1}{\prod_{ij}n_{ij}!}\sum_{\sigma\in S_{n}}\mathrm{Tr}_{\left\{ r\right\} }\left(\Gamma_{R}\left(\sigma^{-1}\right)\right)\prod_{j=1}^{m_{2}}\left(A_{1}B_{2}^{\dagger}\right)_{a_{\sigma^{-1}\left(j\right)}}^{a_{j}}\prod_{i=m_{2}+1}^{n}\left(A_{1}B_{1}^{\dagger}\right)_{a_{\sigma^{-1}\left(i\right)}}^{a_{i}}.
\end{equation}
Therefore, parity changes {$\sigma$} to {$\sigma^{-1}$}. It is worth mentioning that the action of the dilatation operator \eqref{dilatation} on \eqref{parity} produces a similar result to that in \eqref{action}, since we are summing over $\sigma\in S_n$ and the closure of $S_n$ group ensures that for every $\sigma$ there is $\sigma^{-1}\in S_n$.\\
 The action of three-dimensional parity {$P$} takes the {$U(N+l)_k\times U(N)_{-k}$} superconformal theory with {$M-N=l$} to the {$U(N)_k\times U(N+l)_{-k}$} theory \cite{ABJ}. This means that parity flips the levels of the Chern-Simons terms, and consequently produces a different theory. In this case, the action of \eqref{dilatation} on a {$U(N)_k\times U(N+l)_{-k}$} gauge invariant operator of the form \eqref{operator} produces the same result as \eqref{action} except the fact that {$M$} changes to {$N$}.\\
  Let us study the parity operation for the case $m_{2}=2n$ (a similar study for the case with $m_{2}=2n+1$ leads to the same conclusion).
 As we have seen in the previous section, the spectrum of the anomalous dimension for the theory with gauge group {$U(M)_k\times U(N)_{-k}$} is given in \eqref{no-parity1}. The parity operation changes the theory with gauge group {$U(M)_k\times U(N)_{-k}$} to a different theory with gauge group {$U(N)_k\times U(M)_{-k}$} that has the following spectrum
\begin{equation*}
\omega_{i}=8iN\left(\frac{4\pi}{k}\right)^{2},\quad d_{i}=2\left(n-i\right)+1,\quad i=0,1,...,n.
\end{equation*}
Since {$U(M)_k\times U(N)_{-k}$} and {$U(N)_k\times U(M)_{-k}$} Chern-Simons theories are related through the parity transformation, from the spectrum of the dilatation operator of these theories, we observe the following
\begin{equation}\label{commutator1}
\left[D_{non-planar}^{ABJ},\, P\right]\propto\left(M-N\right).
\end{equation}
Thus, the parity operator does not commute with the non-planar ABJ dilatation operator. More precisely, the commutator \eqref{commutator1} is proportional to $M-N$ which indicates the breaking of parity for $M\ne N$. In ABJM theory we have $M=N$ and hence it is parity invariance.  Recall that {$M-N=l$}, then
  \begin{equation*}
  U\left(M\right)_{k}\times\overline{U\left(N\right)}_{-k}\rightarrow U\left(N+l\right)_{k}\times\overline{U\left(N\right)}_{-k},
   \end{equation*}
   \begin{equation*}
   \left[D_{non-planar}^{ABJ},\, P\right]\propto l.
   \end{equation*}
   Similar analysis for the theory with gauge group {$U\left(N\right)_{k}\times\overline{U\left(N+k-l\right)}_{-k}$} gives
\begin{equation}\label{commutator2}
\left[D_{non-planar}^{ABJ},\, P\right]\propto\left(k-l\right).
\end{equation}
From type {$II A$} string theory in \cite{ABJ}, the theory with holonomy {$b_{2}=l/k$} is related to the theory with holonomy {$b_{2}=\left(k-l\right)/k$} by a parity transformation, this corresponds in the field theory side to the equivalence of the {$U\left(N+l\right)_{k}\times\overline{U\left(N\right)}_{-k}$} and {$U\left(N\right)_{k}\times\overline{U\left(N+k-l\right)}_{-k}$} theories. Indeed this is the case in \eqref{commutator1} and \eqref{commutator2} provided that {$\left[D_{non-planar}^{ABJ},\, P\right]\propto {\cal{B}}_{2}$.
%%%%%%%%%%%%%%%%%%%%%%%%%%%%%%%%%%%%%%%%%%%
%%%%%%%%%%%%%%%%%%%%%%%%%%%%%%%%%%%%%%%%%%%
\section{Conclusion}
%%%%%%%%%%%%%%%%%%%%%%%%%%%%%%%%%%%%%%%%%%%
%%%%%%%%%%%%%%%%%%%%%%%%%%%%%%%%%%%%%%%%%%%
We have studied the action of the non-planar dilatation operator for ABJ theory at two loops on operators built from restricted Schur polynomials. The spectrum of the anomalous dimensions signals nonplanar integrability. Our analysis shows that the ABJ theory breaks parity invariance. This is in contrast to the planar two loop dilatation generator which was found to be parity invariant \cite{D.Bak,J.A.Minahan}. When  ABJ theory reduces to ABJM theory, parity invariance is recovered as expected. In this analysis we note that parity breaking does not destroy integrability, this in agreement with the result of \cite{D.Bak,J.A.Minahan}. Furthermore, in the field theory, we have found that {$\left(M-N\right)$} is related to the holonomy {${\cal{B}}_{2}$} of the string theory side. We have considered the case where irreducible representation {$R$} of the gauge invariant operator \eqref{operator} is a Young diagram with two long rows. In this case our operators have a dimension of order {$O(N)$}. We have solved \eqref{spectrum1} within the continuum limit approximation.\\
The exact solution of \eqref{action} in other limits may also reveal whether the ABJ theory remains integrable or not. Another direction to extend this work is to study the action of dilatation operator on operators with dimension of order {$O(N^2)$}. This requires our representation {$R$} to be Young diagram with long rows and long columns. This direction will enrich the AdS/CFT dictionary. However, studying the action of dilatation operators of both {${\cal{N}}=4$} SYM theory and ABJ(M) theories for operators of order {$O(N^2)$} remains a challenge.
\section*{Acknowledgments}
I would like to thank my advisor, professor Robert de Mello Koch for introducing this problem and for enjoyable, helpful discussions. This work is supported by the German Academic Exchange Services DAAD and by the South African Research Chairs Initiative of the Department of Science and Technology and National Research Foundation.
\begin{appendix}
\section{Two point function of ABJ operators in the $SU(2)\times SU(2)$ subsector}
In this appendix, we compute the bare two point function of ABJ operators in the $SU(2)\times SU(2)$ subsector. In this subsector, $n_1$ counts the number of $A_1$s where $n_1=m_1+m_2$, $m_1$ is the number of $B_1$ fields and $m_2$ is the number of $B_2$ fields. Thus, the possible combinations between $A$s and $B$s are $A_1B^{\dagger}_1$ and $A_1B^{\dagger}_2$. the number of pairs $A_1B^{\dagger}_1$ and $A_1B^{\dagger}_2$ are denoted by $n_{11}$ and $n_{12}$ respectively. The gauge invariant operators in terms of restricted Schur polynomials is thus
$$
O_{R,\{r\}}=\frac{1}{n_{11}!n_{12}!}\sum_{\sigma\in S_{n}}\textrm{Tr}_{\{r\}}\left(\Gamma_{R}\left(\sigma\right)\right)\prod_{i=1}^{n}\left(A_{1}\right)_{\alpha_{i}}^{a_{i}}\prod_{i=1}^{m_{1}}\left(B_{1}^{\dagger}\right)_{a_{\sigma\left(i\right)}}^{\alpha_{i}}\prod_{j=1+m_{1}}^{n}\left(B_{2}^{\dagger}\right)_{a_{\sigma\left(j\right)}}^{\alpha_{j}}
$$
$$
\equiv\textrm{Tr}\left(P_{R,\{r\}}A_{1}^{\otimes n_{11}+n_{12}}\left(B_{1}^{\dagger}\right)^{\otimes n_{11}}\left(B_{2}^{\dagger}\right)^{\otimes n_{12}}\right),
$$
where the projector $P_{R,\{r\}}$ is given by
$$
P_{R,\{r\}}=\frac{1}{n_{11}!n_{12}!}\sum_{\sigma\in S_{n}}\textrm{Tr}_{\{r\}}\left(\Gamma_{R}\left(\sigma\right)\right)\sigma
$$
In the above expression, we note that $n_{11}+n_{12}=n$, $n_{11}=m_1$ and $n_{12}=m_2$. In this case the two point function can be written as
$$
\bigl\langle O_{R,\{r\}}O_{S,\{s\}}^{\dagger}\bigr\rangle=\sum_{\sigma\in S_{n}}\sum_{\rho\in S_{m_{1}}\times S_{m_{2}}}\textrm{Tr}\left(P_{R,\{r\}}\sigma P_{S,\{s\}}\rho\right)\textrm{Tr}\left(\sigma^{-1}\rho^{-1}\right)
$$
The sum over $\rho$ gives $m_1!m_2!$. Therefore
$$
\bigl\langle O_{R,\{r\}}O_{S,\{s\}}^{\dagger}\bigr\rangle=m_{1}!m_{2}!\sum_{\sigma\in S_{n}}\textrm{Tr}\left(P_{R,\{r\}}\sigma P_{S,\{s\}}\right)\textrm{Tr}\left(\sigma^{-1}\right)
$$
Using the identity (proved in \cite{Schur-polynomials1})
\begin{equation}\label{identity0}
P_{R,\{r\}}P_{S,\{s\}}=\delta_{RS}\delta_{\{r\},\{s\}}\frac{n!}{n_{11}!n_{12}!}P_{R,\{r\}},
\end{equation}
we get
\begin{equation}\label{two-point1}
\bigl\langle O_{R,\{r\}}O_{S,\{s\}}^{\dagger}\bigr\rangle=\frac{n!}{d_{R}}\delta_{RS}\delta_{\{r\},\{s\}}\sum_{\sigma\in S_{n}}\textrm{Tr}\left(P_{R,\{r\}}\sigma\right)\textrm{Tr}\left(\sigma^{-1}\right).
\end{equation}
From \cite{nonplanar-ABJM}, the trace $\textrm{Tr}\left(\sigma^{-1}\right)$ can be written as
$$
\textrm{Tr}\left(\sigma^{-1}\right)=\sum_{T\vdash n}\chi_{T}\left(\sigma^{-1}\right)Dim\left(T\right),
$$
where $Dim\left(T\right)$ is the multiplicity factor and it is defined by
\begin{equation}\label{multplicity}
Dim\left(T\right)=\frac{f_{T}\left(M\right)}{\textrm{hooks}_{T}}.
\end{equation}
$f_T(M)$ is the product of the box weight in the irrep $T$,  $\textrm{hooks}_{T}$ is the product of hooks of each individual box in the irrep $T$ (for examples of computing the hooks and $f_T$,  see \cite{Koch:2007uu,Schur-polynomials1,Schur-polynomials2}). In \eqref{multplicity} we get $f_T(M)$ instead of $f_T(N)$ due to $\sigma^{-1}$ which control the contractions among the Greek indices where $\alpha,\beta=1,...,M$. \eqref{two-point1} becomes
\begin{equation}\label{two-point2}
\bigl\langle O_{R,\{r\}}O_{S,\{s\}}^{\dagger}\bigr\rangle=\frac{n!}{d_{R}}\delta_{RS}\delta_{\{r\},\{s\}}\sum_{T\vdash n}\frac{f_{T}\left(N\right)}{\textrm{hooks}_{T}}\sum_{\sigma\in S_{n}}\textrm{Tr}\left(P_{R,\{r\}}\sigma\right)\chi_{T}\left(\sigma^{-1}\right).
\end{equation}
Identify
$$
\frac{n!}{d_{T}}P_{T,\{t\}}=\sum_{\sigma\in S_{n}}\chi_{T}\left(\sigma^{-1}\right)\sigma^{-1}
$$
Thus \eqref{two-point2} becomes
\begin{align}
\bigl\langle O_{R,\{r\}}O_{S,\{s\}}^{\dagger}\bigr\rangle=&\frac{n!}{d_{R}}\delta_{RS}\delta_{\{r\},\{s\}}\sum_{T\vdash n}\frac{f_{T}\left(N\right)}{\textrm{hooks}_{T}}\frac{n!}{d_{T}}\textrm{Tr}\left(P_{R,\{r\}}P_{T,\{t\}}\right)\nonumber\\
=&\frac{f_{R}\left(M\right)f_{R}\left(N\right)n!}{d_{R}}\delta_{RS}\delta_{\{r\},\{s\}}\textrm{Tr}\left(P_{R,\{r\}}\right).\nonumber
\end{align}
The value of $\textrm{Tr}\left(P_{R,\{r\}}\right)$ is \cite{Koch:2007uu}
$$
\textrm{Tr}\left(P_{R,\{r\}}\right)=\frac{\textrm{hooks}_{R}d_{R}}{\textrm{hooks}_{r_{11}}\textrm{hooks}_{r_{12}}n!}
$$
Thus, the bare two pint function in the $SU(2)\times SU(2)$ subsector of ABJ theory is
\begin{equation}
\bigl\langle O_{R,\{r\}}O_{S,\{s\}}^{\dagger}\bigr\rangle=\delta_{RS}\delta_{\{r\},\{s\}}\frac{\textrm{hooks}_{R}f_{R}\left(M\right)f_{R}\left(N\right)}{\textrm{hooks}_{r_{11}}\textrm{hooks}_{r_{12}}}.
\end{equation}

\end{appendix}

\end{document}